\documentclass[12pt,preprint]{aastex}

\slugcomment{Submitted to the Astrophysical Journal Letters}

\shorttitle{An M8.5 Dwarf With Proper Motion
$\mu=2.38\arcsec$ yr$^{-1}$}
\shortauthors{Lepine, Rich, \& Shara}

\begin{document}

\title{Discovery of an M8.5 Dwarf With Proper Motion
$\mu=2.38\arcsec$ yr$^{-1}$}

\author{S\'ebastien L\'epine\altaffilmark{1}, R. Michael
Rich\altaffilmark{2,3}, James D. Neill\altaffilmark{1}, Adeline
Caulet\altaffilmark{4}, \and Michael M. Shara\altaffilmark{1}}
\altaffiltext{1}{Department of Astrophysics, Division of Physical
Sciences, American Museum of Natural History, Central Park West at
79th Street, New York, NY 10024, USA}
\altaffiltext{2}{Visiting Astronomer, Lick Observatory}
\altaffiltext{3}{Department of Physics and Astronomy, University of
California at Los Angeles, Los Angeles, CA 90095, USA}
\altaffiltext{4}{Calypso Observatory, Kitt Peak, Arizona}

\begin{abstract}
We report the discovery of LSR1826+3014, a very faint
(V=19.36) star with a very large proper motion ($\mu=2.38\arcsec$
yr$^{-1}$). A low resolution red spectrum reveals that LSR1826+3014 is
an ultra-cool red dwarf with spectral type M8.5 V and with a radial
velocity $v_{rad}=+77\pm10$km s$^{-1}$. LSR1826+3014 is thus the
faintest red dwarf ever discovered with a proper motion larger than
$2\arcsec$ yr$^{-1}$. Optical and infrared photometry suggest that the
star is at a distance $d=13.9\pm3.5$pc from the Sun, which implies it
is moving relative to the local standard of rest with a total velocity
of $175\pm25$km s$^{-1}$. Numerical integration of its orbit suggests
that LSR1826+3014 is on a halo-like galactic orbit.
\end{abstract}


\section{Introduction}

Very few stars are known to have proper motions $\mu>2\arcsec$
yr$^{-1}$. The LHS catalog \citep{L79} lists only 60 systems
(which include 9 resolved doubles and 2 resolved triples) with proper
motions in that range, and few systems have been added since. The
discovery of a faint $\mu=2.13\arcsec$ yr$^{-1}$ white dwarf
(\objectname[]{ER8}=\objectname[]{GJ3770}) was reported by
\cite{RMWG86}. More recently, the candidate halo white dwarf
\objectname[]{F351-50} was discovered by \citet{IIBSG00}; its proper
motion is $\mu=2.33\arcsec$ yr$^{-1}$.

Most of these extremely rare objects are very nearby stars, within
10pc of the Sun. \objectname[]{Proxima}, our closest neighbor, has a
proper motion $\mu=3.853 \arcsec$ yr$^{-1}$ \citep{E97}. However, a
few of the very high proper motion stars lie beyond 10pc, which means
they have velocities relative to the Sun that are larger than 100 km
s$^{-1}$, and they are likely to be local members of the galactic
thick disk or halo. One example is the nearby subdwarf
\objectname[]{LHS 42} ($\mu=3.21 \arcsec$ yr$^{-1}$, spectral type
sdK4), which is a very likely member of the galactic halo
\citep{FJ98}. The star with the largest known proper motion is the
star discovered by \citet{B16}, which bears his name and is moving at
$\mu=10.3\arcsec$ yr$^{-1}$; Barnard's star, with spectral type M4.0
V \citep{KHM91}, is believed to be a member of the thick disk
\citep{AM87}.

Apart from the two faint white dwarfs mentioned above, all the
known stars with $\mu>2\arcsec$ yr$^{-1}$ are relatively bright;
none of them is fainter than R=15. This may be an observational
bias, since faint stars with very large proper motions are notoriously
difficult to find. On the other hand, it is the faintest components of
the halo that are expected to be detected as faint high proper motion
stars. The fact that few faint stars with large proper motion are
known may simply reflect the very low local density of low-luminosity
halo stars. Clearly, the discovery of faint ($R>15$) stars with proper
motions $\mu>2\arcsec$ yr$^{-1}$ can have profound implications for our
understanding of the density, composition, and kinematics of the halo.

In this Letter, we report the discovery of
\objectname[]{LSR1826+3014}, a very faint (R$>17$) star with a proper
motion of $\mu=2.38\arcsec$ yr$^{-1}$. Our spectroscopy shows the star
to be an M8.5 dwarf at a distance of about 13.9pc, moving relative to
the local standard of rest with a velocity of about 175 km
s$^{-1}$. This low-mass star appears to have an orbit that is most
consistent with membership in the galactic halo.

\section{Proper Motion Discovery}

The very high proper motion star \objectname[]{LSR1826+3014} was
discovered through our automated search for stars with large proper
motion using the Digitized Sky Survey \citep{LSR02}, undertaken as
part of the NStars Program. The star was found in a relatively
crowded Milky Way field at galactic latitude b=+18.31. The very
large proper motion of the star is above the detection threshold of
our SUPERBLINK software, and it was not initially flagged as a high
proper motion star. However, SUPERBLINK also identifies candidate
variable objects as it compares the Digitized Sky Survey POSS-I
(xx103aE+plexi) and POSS-II (IIIaF+RG610) scanned red
plates. \objectname[]{LSR1826+3014} was first noticed as a pair of
candidate variable stars lying within about $2\arcmin$ of each other;
one star showing up in the POSS-I plate only, the other one in the
POSS-II plate only. Suspecting this could be a star with an extremely
large proper motion, we searched for archival images of this field at
different epochs. The Digitized Sky Survey includes 3 more images of
the field around \objectname[]{LSR1826+3014}. The Quick-V Northern
plate (IIaD+W12) doesn't go deep enough to detect the star; the
POSS-II Blue plate (IIIaJ+GG385) shows a very faint star, near the
detection limit, at the location expected for a high proper motion
star. The POSS-II Near-IR (IVN+RG9) plate shows a relatively bright
star again at the location expected for a star with a very large
proper motion. To provide definitive evidence that the star is a high
proper motion object, we obtained an R band image of the same field at
the Calypso telescope (Kitt Peak) in April 2002. We found again the
star at the exact location expected for a very high proper motion
object.

The 3 epochs of red band images are displayed in Figure 1. Shown are
$4\arcmin \times 4\arcmin$ sections of the POSS-I red, POSS-II red,
and Calypso R-band image. The high proper motion star is indicated by
a circle. The three images span over 50 years, during which time
\objectname[]{LSR1826+3014} has moved by a little more than
$2\arcmin$. 

We reprocessed the POSS-I and POSS-II plates for this region of the
sky with SUPERBLINK, this time allowing for the detection of stars
with proper motions up to $3\arcsec$ yr$^{-1}$. This time the software
easily identified \objectname[]{LSR1826+3014}, and calculated a proper
motion $\mu=2.38\arcsec$ yr$^{-1}$. The proper motion measured by
SUPERBLINK is the proper motion of the star calculated relative to the
mean position of all objects within $4\arcmin$ of the candidate high
proper motion star (image superposition); the astrometric solutions
from the first Digitized Sky Survey are used to obtain the local scale
of the scanned plate.

With a proper motion $\mu=2.38\arcsec$ yr$^{-1}$,
\objectname[]{LSR1826+3014} ranks as the stellar system with the 42nd
largest proper motion on the sky, and one of the faintest stars known
with a proper motion larger than 2$\arcsec$
yr$^{-1}$. \objectname[]{LSR1826+3014} appears to be headed towards
the bulge of a background Galaxy (see Figure 1). By extrapolating the
motion of the star, we estimate that it will move within $0.4\arcsec$
of the centroid of the core of that galaxy around the 2006.5 epoch.

\section{Photometry}

We observed \objectname[]{LSR1826+3014} on 12 April, 2002 under
photometric conditions using the Calypso Observatory \citep{NSCB02} Wide
Field Camera in the Johnson BV, and Cousins RI filters. Standard stars
from the \citet{L92} catalog were observed and used for the absolute
calibration of the object. The Landolt magnitudes of the standard
stars were converted to the Johnson-Cousins system. All frames were
debiased and flat-fielded using IRAF \citep{T86}. The object and
standard stars were measured in each of multiple frames using the
apphot package in IRAF and the results were averaged.  We used
apertures with FWHM comparable to the seeing (4.5 pixels =
0.675$\arcsec$) and used bright stars to derive a correction to 100\%
light. Using the multiple frames we derived an external error for the
measurements of 0.03 mag in the B filter and 0.01 mag or better in the
other filters. The final B, V, R and I magnitudes for the star are
listed in Table 1.

The star \objectname[]{LSR1826+3014} was also identified in the 2MASS
second incremental release as the bright point source
\objectname[]{2MASS1826113+301420}. The corresponding 2MASS source was
found within 0.33$\arcsec$ of the predicted location of
\objectname[]{LSR1826+3014} at the epoch of the 2MASS images. The
2MASS J, H, and K$_s$ magnitudes are listed in Table 1.

\section{Spectroscopy}

The star was observed at Lick observatory on July 6, 2002
using the Kast spectrograph installed at the Cassegrain focus of the 3m
Shane telescope. We used the 600 l/mm grating blazed at 7500\AA\ to
obtain a spectrum covering the range 6300\AA-9100\AA\ with a resolution
of 2.33\AA\ per pixel. The star was imaged through a $2.3\arcsec$ wide
slit, and with the slit oriented vertically to avoid slit loss due to
differential atmospheric refraction. Standard spectral reduction was
performed with the specred package in IRAF, including normalization
using a spectrophotometric standard and removal of telluric lines. The
resulting spectrum is displayed in Figure 2.

Spectral classification was performed by visual comparison with the
standard sequence found in \citet{KHM91}. We found our spectrum of
\objectname[]{LSR1826+3014} to be most similar to an M8.5 V dwarf. We
also used the values of the standard M dwarf spectral indexes defined
in \citet{LRS02}. The combined values of the VO1, TiO6, VO2, and TiO7
indexes are consistent with a spectral type M8.5 V.

The radial velocity was estimated by comparing the spectrum of
\objectname[]{LSR1826+3014} with the spectrum of the radial velocity
standard HR7002, an M6 giant, which we had observed immediately after
\objectname[]{LSR1826+3014} on the night of July 6. We estimated the
difference in radial velocity between the two stars by shifting the
spectrum of HR7002 to match the TiO bandhead at 7050\AA\ (the two stars
had virtually the same heliocentric correction at the time of
observation). The best fit was obtained with a +6 km s$^{-1}$ shift of
the HR7002 template, with a possible error of $\pm$ 10 km
s$^{-1}$. The star HR7002 has a measured radial velocity
v$_{rad}$=+70.8 km s$^{-1}$ \citep{FPTWWS94} which yields a radial
velocity v$_{rad}$=$+77\pm10$ km s$^{-1}$ for
\objectname[]{LSR1826+3014}.

\section{Distance and kinematics}

We use the newly calibrated ($M_V,V-K$) and ($M_I,J-K$) relationships
of \citet{RC02} to obtain a photometric parallax for
\objectname[]{LSR1826+3014}. The $V-K=8.56$ color implies an absolute
magnitude $M_V$=18.72 and suggests a distance of d=$13.5$pc. The
$I-J=2.69$ color implies an absolute magnitude $M_I$=13.47, which in
turn suggests a distance of d=$15.0$pc. We also use the absolute
magnitude / spectral type relationship calibrated by \citet{LRS02} to
obtain a spectroscopic parallax for the star; a spectral type of M8.5V
is consistent with an absolute magnitude $M_{K_s}$=10.18, which
suggests a distance of d=$13.2$pc. Note that the spectroscopic
distance is consistent with the photometric distance.

The mean value of the three distance estimates is d=$13.9$pc. Each of
the color/magnitude and spectral type/magnitude relationships has a
scatter of about $\pm0.5$ magnitudes, which corresponds to a distance
error of $\pm$25\%, or $\pm$3.5pc. We therefore adopt for
\objectname[]{LSR1826+3014} an estimated distance of
d$=13.9\pm3.5$pc. These results can be compared with the recent
compilation of late-type M dwarfs and L dwarfs with geometric
parallaxes by \citet{D02}. One finds that the $I$, $J$, and $K_s$
magnitudes of \objectname[]{LSR1826+3014} are all consistent with an
M8.5 dwarf at a distance of 13-15pc.

Using the proper motion, distance estimate, and measured radial
velocity for \objectname[]{LSR1826+3014}, we calculate the motion
of the star relative to the local standard of rest in $[U,V,W]$ space,
where $U$ is the velocity toward the galactic center, $V$ is the
velocity toward the direction of galactic rotation, and $W$ the
velocity toward the north galactic pole. We use as the motion of the
Sun relative to the local standard of rest the values estimated by
\citet{DB98}: $[U,V,W]=[+10,+5,+7]$km s$^{-1}$. We find the velocity
components of \objectname[]{LSR1826+3014} to be $U$=+92$\pm$22km
s$^{-1}$, $V$=-26$\pm$17km s$^{-1}$, and $W$=+134$\pm$30 km
s$^{-1}$ relative to the local standard of rest.

Using these values, we can calculate the probable orbit of the
star. We use the galactic mass model by \citet{DC95}, which includes
separate terms for the bulge, disk, and halo. We integrate with a
Runge-Kutta fourth order integrator, in time steps of $10^3$
yr. Figure 3 shows a 950 Myr integration of the orbit of the star
plotted in the [R,z] plane, where R is the galactocentric distance in
cylindrical coordinates, and z is the distance from the plane. The
star is shown to oscillate between 5.5Kpc$<$R$<$14.5Kpc, and
-6.0Kpc$<$z$<$+6.0Kpc, which strongly suggests that
\objectname[]{LSR1826+3014} is a member of the galactic halo. Its
spectrum, however, is not consistent with a very metal-poor object. It
is thus possible that \objectname[]{LSR1826+3014} is a star initially
born in the disk that later got ejected into the halo.

Based on its kinematics, \objectname[]{LSR1826+3014} is most likely to
be a relatively old ($\geq$5Gyr) object. Comparison with the
evolutionary model of \citet{CBAH00} shows that the J and K magnitudes
of \objectname[]{LSR1826+3014} are consistent with those of a 5-10 Gyr
old star with a mass $0.08M_{\sun}<M_*<0.09M_{\sun}$. The star is thus
relatively close to the hydrogen burning limit.

\section{Conclusions}

We have discovered one of the faintest stars on the sky with a proper
motion $>2\arcsec$yr$^{-1}$. The star \objectname[]{LSR1826+3014} is a
low mass red dwarf of spectral type M8.5V at a distance of
13.9$\pm$3.5pc from the sun. It is moving on the sky with a proper
motion of 2.38$\arcsec$ yr$^{-1}$ and has a spectroscopic radial
velocity $v_{rad}$=+77$\pm$10 km s$^{-1}$. The star has a very large
motion relative to the local standard of rest. Its calculated galactic
orbital motion strongly suggests that it is a member of the halo,
reaching heights of about 6 Kpc above the plane of the galactic disk. 

A trigonometric parallax should be obtained to confirm the extreme
value of its motion relative to the local standard of rest. Although
the star does not appear to be significantly metal poor, its status as
a possible disk star ejected from the plane, an extreme old disk star,
or a true halo star should be addressed by careful measurement of its
metal abundance. That an old hydrogen burning object should be so cool
as to have the spectrum of an M8.5 dwarf indicates that it must be
extremely close to the hydrogen burning limit. That makes
\objectname[]{LSR1826+3014} a unique object among the high velocity
stars.

\acknowledgments

This research program is being supported by NSF grant AST-0087313 at
the American Museum of Natural History, as part of the NStars Program.

This publication makes use of data products from the Two Micron All
Sky Survey, which is a joint project of the University of
Massachusetts and the Infrared Processing and Analysis Center, funded
by the National Aeronautics and Space Administration and the National
Science Foundation.

\newpage

\begin{deluxetable}{lrl}
\tabletypesize{\scriptsize}
\tablecolumns{3} 
\tablewidth{0pt} 
\tablecaption{Basic Data for LSR1826+3014} 
\tablehead{Datum & Value & Units}
\startdata 
RA (2000.0) & 18:26:11.03 & h:m:s\\
DEC (2000.0) & +30:14:19.1 & d:m:s\\
$\mu$       & 2.38 & $\arcsec$ yr$^{-1}$\\
pma         & 253.3 & $\degr$\\
v$_{rad}$   & 77$\pm$10 & km s$^{-1}$\\
B & 21.46$\pm$0.03 & mag\\
V & 19.36$\pm$0.01 & mag\\
R & 17.40$\pm$0.01 & mag\\
I & 14.35$\pm$0.01 & mag\\
J & 11.66$\pm$0.03 & mag\\
H & 11.16$\pm$0.03 & mag\\
K$_s$ & 10.80$\pm$0.04 & mag\\
Spectral Type & M8.5 V & \\
Distance & 13.9$\pm$3.5 & pc \\
$U$ & +92$\pm$22 & km s$^{-1}$\\
$V$ & -26$\pm$17 & km s$^{-1}$\\
$W$ & +134$\pm$30 & km s$^{-1}$
\enddata
\end{deluxetable} 





\newpage

\begin{figure}
\plotone{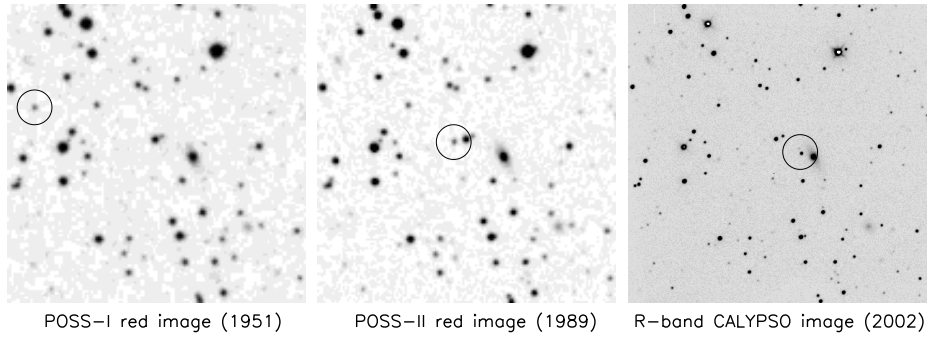}
\caption{\label{fig1} The new high proper motion star
LSR1826+3014. Left: red plate of the first epoch Palomar Sky Survey,
obtained in 1951. Middle: red plate of the second epoch Palomar Sky
Survey, obtained in 1989. Right: R band Calypso image, obtained in
2002. All the fields are $4.0\arcmin$ on the side, with north up and
east left. Circles are drawn centered on the location of LSR1826+3014
at each epoch. LSR1826+3014 appears to be headed toward the core of a
background galaxy.}
\end{figure}

\begin{figure}
\plotone{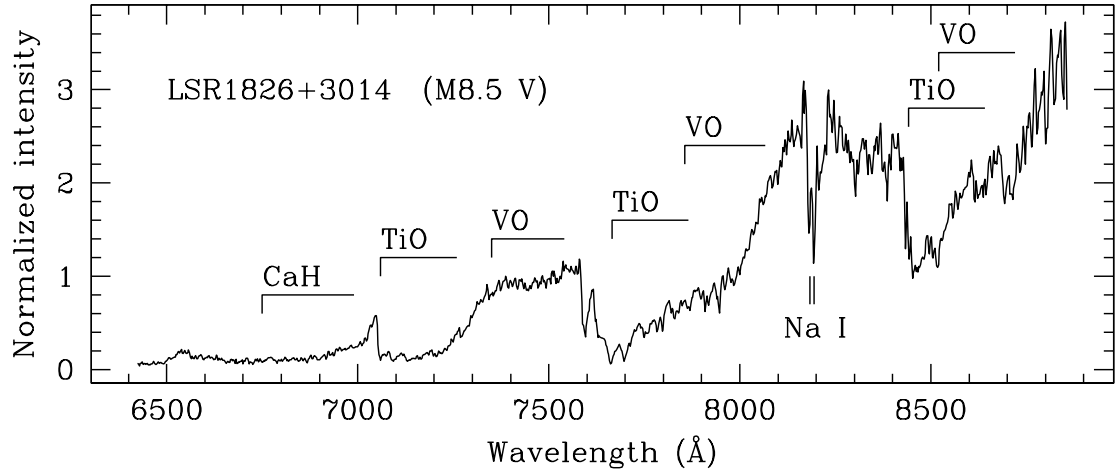}
\caption{\label{fig2} Optical spectrum of the high proper motion
star LSR1826+3014 obtained with the Kast spectrograph on the 3m Shane
Telescope at Lick Observatory. The most prominent spectral features
are identified. The spectrum is most consistent with a spectral type
M8.5 V.}
\end{figure}

\begin{figure}
\plotone{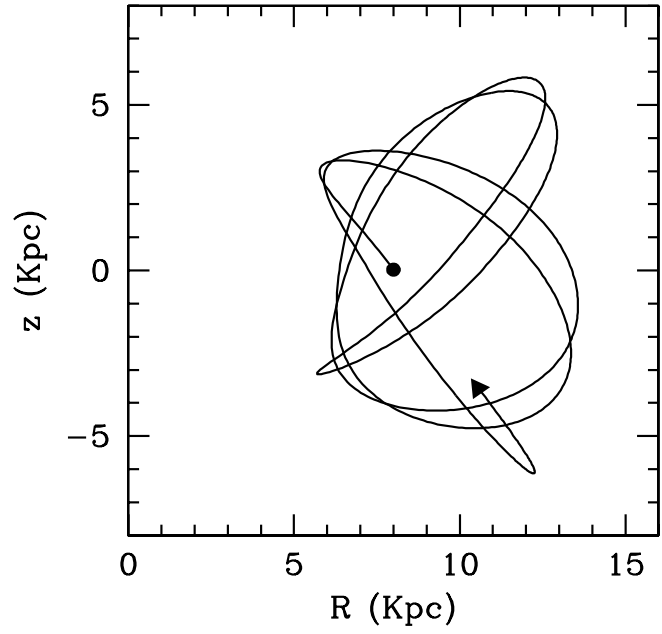}
\caption{\label{fig3}Galactic orbital motion of LSR1826+3014
extrapolated over the next 950 Myr (dot: starting point, arrowhead:
ending point). The star's trajectory takes it up to 6 Kpc from the
plane of the disk.}
\end{figure}

\end{document}